\documentclass{article}
\usepackage{spconf}
\usepackage{url}
\usepackage{cite}
\usepackage{amsmath,amssymb,amsfonts}
\usepackage{algorithmic}
\usepackage{graphicx}
\usepackage{textcomp}
\usepackage{xcolor}

\usepackage{listings}
\usepackage{caption}
\usepackage{subcaption}
\usepackage[ruled,linesnumbered,vlined]{algorithm2e}
\usepackage{hyperref}
\hypersetup{colorlinks,allcolors=black}

\newcommand*{\cdbackslash}{\_\kern-.1355ex\_}
\title{Cryo-RALib - a modular library for accelerating alignment in cryo-EM}

\name{Szu-Chi Chung$^{\star}$, Cheng-Yu Hung$^{\star}$, Huei-Lun Siao$^{\star}$, Hung-Yi Wu$^{\star}$, Wei-Hau Chang$^{\dagger}$, I-Ping Tu$^{\star}$}
\address{$^{\star}$Institute of Statistical Science, Academia Sinica. \\ 
$^{\dagger}$Institute of Chemistry, Academia Sinica. }

\begin{document}

%
%
%

\maketitle

%
%

\begin{abstract}
Thanks to automated cryo-EM and GPU-accelerated processing, single-particle cryo-EM has become a rapid structure determination method that permits capture of dynamical structures of molecules in solution, which has been recently demonstrated by the determination of COVID-19 spike protein in March, shortly after its breakout in late January 2020. This rapidity is critical for vaccine development in response to emerging pandemic. This explains why a 2D classification approach based on multi-reference alignment (MRA) is not as popular as the Bayesian-based approach despite that the former has advantage in differentiating structural variations under low signal-to-noise ratio. This is perhaps because that MRA is a time-consuming process and a modular GPU-acceleration library for MRA is lacking. Here, we introduce a library called \textit{Cryo-RALib} that expands the functionality of CUDA library used by GPU ISAC. It contains a GPU-accelerated MRA routine for accelerating MRA-based classification algorithms. In addition, we connect the cryo-EM image analysis with the python data science stack so as to make it easier for users to perform data analysis and visualization. Benchmarking on the TaiWan Computing Cloud (TWCC) container shows that our implementation can accelerate the computation by one order of magnitude. The library is available at 
\href{https://github.com/phonchi/Cryo-RAlib}{https://github.com/phonchi/Cryo-RAlib}.

\end{abstract}


\section{Introduction}
In contrast to X-ray crystallography, cryo-EM is a method that is amenable to the structural determination of proteins in non-crystalline state. With the instrument automation and advances in algorithms, single particle cryo-EM has become a mainstream tool to solve 3D structures of molecules at near-atomic resolution. It is noted that as the automated data collection is becoming mature, the equipment that is up-running 24/7 can generate 1000 to 4000 micrographs per day. Therefore, the GPU hardware has been invoked to accelerate the cryo-EM workflow to meet the demand for processing a large volume of data. The GPU is now used in various steps including movie alignment \cite{zheng2017motioncor2}, contrast transfer function estimation  \cite{zhang2016gctf}, identification of particles within micrographs \cite{wagner2019sphire}, Bayesian 2D classification algorithm like RELION \cite{Relion} or cryoSPARC \cite{punjani2017cryosparc} and 3D refinement algorithms \cite{Relion,punjani2017cryosparc}. Even with GPU acceleration, the popular RELION 2D classification usually takes several days to finish the classification task. Evidently, image classification has become the bottleneck in the workflow. We and others notice that the computational complexity of the Bayesian approach is much higher than the one based on multireference alignment (MRA).  The complexity is at least in the order of $O(nL^4 \log L)$  compared with $O(nL^3)$ of MRA, where $n$ is the sample size and $L$ is the pixel number for one direction of the particle image, and $t$ is the translational shift search range \cite{singer2020computational}. 
Besides, to further extend to atomic resolution, one must carefully deal with heterogeneity within the image data. When such is concerned, an MRA-based method has advantage because it can better differentiate subtle structure differences under low SNR compared with the Bayesian approach \cite{wu2017massively, chung2020pre}.


\section{Previous Works}

2D and 3D classification algorithms based on the MRA or Bayesian approach are standard steps in cryo-EM workflow \cite{singer2020computational}. The 2D algorithm is depicted as follows. We use 2D to simplify the illustration while the extension to 3D is straightforward. First, K initial 2D average images are randomly generated or provided by the user. Second, the particle images are compared with those K initials as references by tuning parameters of shifts and in-plane rotations. The reference that maximizes the cross-correlation (CC) is to be recorded for each image. The new 2D averages can then be updated with the aligned images. After multiple iterations of refinement on parameters, the class assignment of each image along with its shift and rotation parameters to a particular 2D class is usually unambiguous. The popular Bayesian approach like RELION \cite{Relion}  is slightly different in that it uses fuzzy assignments for both alignment parameters and classes instead of the best one. The 2D averages are then obtained through weighted averages over all possible orientations and classes.
Although the Bayesian approach is very powerful in 3D refinement \cite{Relion}, the algorithms based on MRA have been shown to provide better performance in 2D classification. For instance, CL2D \cite{sorzano2010clustering} can reduce the influence of noise and avoid the unbalance class phenomenon, ISAC \cite{yang2012iterative} is an algorithm that ensures robust classes to be output by repeated tests, while Prime \cite{reboul2016stochastic} can escape the local minimum by adopting stochastic hill-climbing.  There are many implementation strategies for the alignment step in MRA \cite{joyeux2002efficiency}. Among these implementations, re-sample to polar coordinate (RPC) has been shown to have the best performance under low SNR so that we use it in this work. It is noted that most of the works do not exploit GPU in the MRA step. However, since the calculation of similarities is independent of particles, 2D reference and orientation, it is possible to calculate them in parallel using GPU. 

\begin{figure}[htb]
\centering
\includegraphics[width=3.2in]{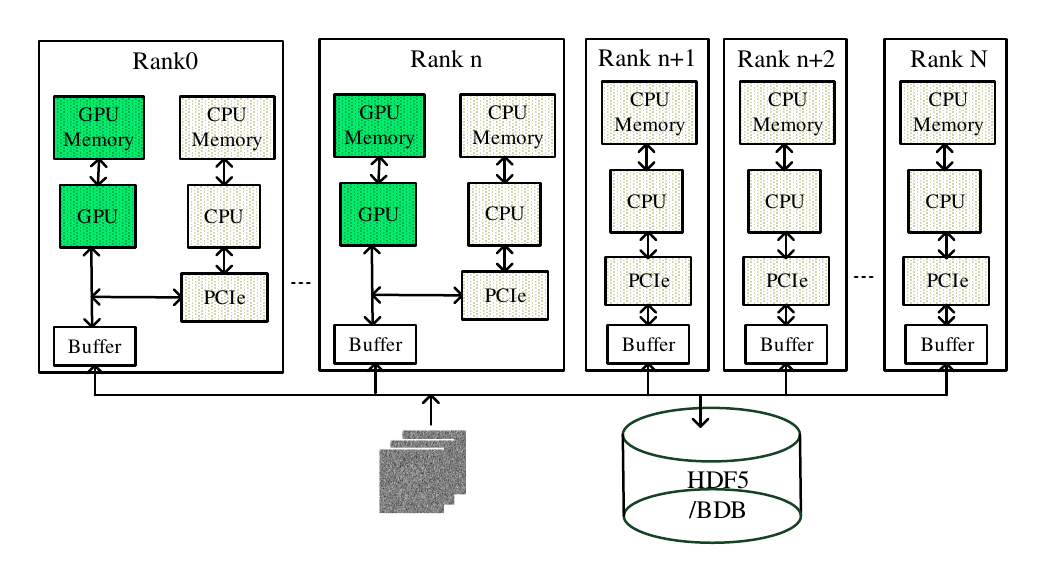} 
\caption{The architecture is a summary of the coding framework underlying the GPU-accelerated ISAC \cite{GPU-ISAC}. We employ this architecture to implement MRA.}
\label{fig:framework}
\end{figure}

Fortunately, RPC has been GPU-accelerated in reference-free alignment (RFA) of GPU ISAC \cite{GPU-ISAC}, whose code has been released in July 2020. We summarize the architecture of the coding framework underlying the GPU-accelerated ISAC in Fig. \ref{fig:framework}.
This architecture is implemented using message passing interface (MPI) and Nvidia CUDA and the processing flow is described as follows. First, the particle images are read in parallel and immediately pre-processed by all $N$ MPI processes. Second, every process is sent to the first $n$ MPI processes that contain GPU resources through MPI. 
Third, the first $n$ MPI processes perform the alignment algorithm. Finally, the first $n$ MPI processes send the data back to all $N$ MPI processes and all the processes write the metadata and the transformed images to the disk. Notice that the main task in RPC procedure is to compute the rotation cross-correlation function. The function is  defined as $c(\phi) = \int_{r_1}^{r_2}\int_{0}^{2\pi} x(r, \theta) y(r, \theta+\phi)|r| \,d\theta \,dr$. 
To speed up the host-to-device memory transfer, the authors in \cite{GPU-ISAC} employ the texture memory to store the image as well as its metadata, and a floating array in global memory is used to store the images after polar conversion, Fast Fourier Transform (FFT), Inverse FFT (IFFT) or alignment. The results of the CC computation are stored in a table which holds the CC information of all (mirrored) images with all reference images together with all shifts. Each row in the table stores all CC information for one image.

\section{The MRA Framework}
In this work, we exploit the parallelism and architecture mentioned in Section 2. Our framework is built upon the general-purpose cryo-EM processing library called EMAN2 \cite{tang2007eman2} and the MRA is based on GPU ISAC \cite{GPU-ISAC}. The architecture and data layout of MRA we use is from the GPU-accelerated RFA in \cite{GPU-ISAC}. We describe the acceleration of MRA in this section and provide a simple link between cryo-EM image processing and the python data science stack to enable rapid development of GPU-accelerated operations which is described in Section 4. The programs are integrated into a library called \textit{Cryo-RALib} which is accessible to the cryo-EM community.  

\begin{algorithm}[tbph]
\caption{GPU Accelerated MRA}
\label{alg:GMRA}
\SetKwBlock{DoParallel}{do in parallel}{end}
\SetKwInOut{Input}{Input}  
\Input{Set of particle images \textit{$X$} \\
Set of reference images \textit{$Y$}  
}
\KwOut{Set of class averages $Y'$}  
 
 
 
 \Repeat{convergence}{
 
 \For{each batch}{
 Transfer $X$ and $Y$ in this batch from host memory into texture memory in device.

\DoParallel{
 Convert $y_i$ to polar coordinates and apply FFT to obtain $y'_i$.
 }
 \For {each shift} {
 \DoParallel{
 Convert $x_i$ to polar coordinates with given shift and apply FFT to obtain $x'_i$. Compute CC between $x'_i$ and $y'_i$ and store CC into the table.
 }
 Apply inverse FFT to CC table.
 }
 \DoParallel{
 Find the largest CC value for each image using reduction.
 Find the corresponding alignment parameters and apply them to each image.
 }
 Compute new $Y'$ from aligned images.

 Transfer averages and parameters to host.
 }
 }

 Return $Y'$ and parameters.
\end{algorithm}

Our GPU implementation follows the framework of CPU implementation from EMAN2, as shown in Algorithm 1. In other words, we provide a GPU version of MRA from EMAN2. 
To do so, we made use of existing libraries with optimized CUDA primitives, the polar coordinate conversion, FFT, IFFT and applying alignment parameters are from GPU ISAC \cite{GPU-ISAC}. Our CC computation of $FFT(x)' \times FFT(y)$ in MRA is similar to the one used in RFA of \cite{GPU-ISAC}. The difference is that each block, which processes one particle image, now computes its CC values with all the given reference images instead of single reference. This is done by adding another layer of parallelism for the reference images in the y-dimension of GPU grid. 




\begin{figure}[htb]
\centering
\includegraphics[width=2.6in]{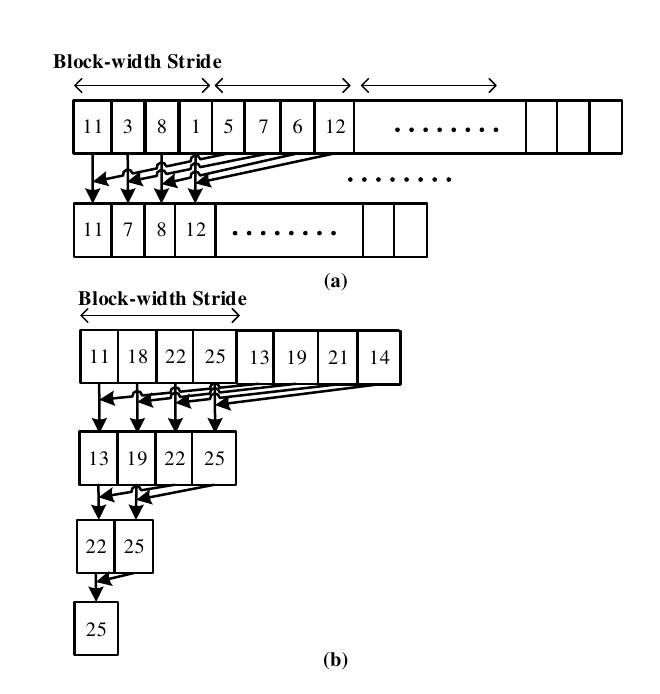} 
\caption{The reduction strategy. (a) Initialize the shared memory. (b) General reduction on the shared memory.}
\label{fig:reduction}
\end{figure}

\begin{figure}[htb]
\centering
\includegraphics[width=2.6in]{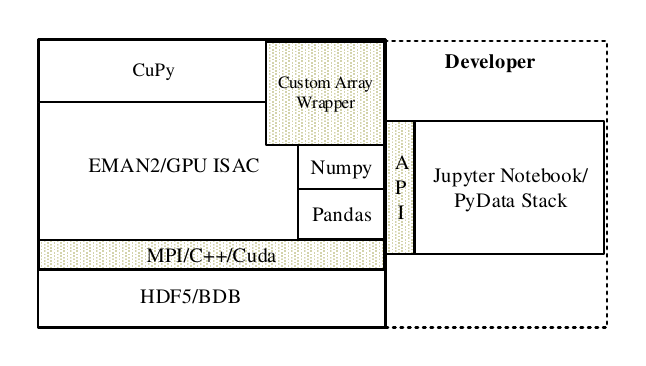} 
\caption{The software stack of our framework. The shaded blocks represent the interfaces with other libraries.}
\label{fig:soft_stack}
\end{figure}

\begin{figure}[htb]
     \centering
     \mbox{
     \begin{subfigure}[b]{0.20\textwidth}
         \centering
         \includegraphics[width=\textwidth]{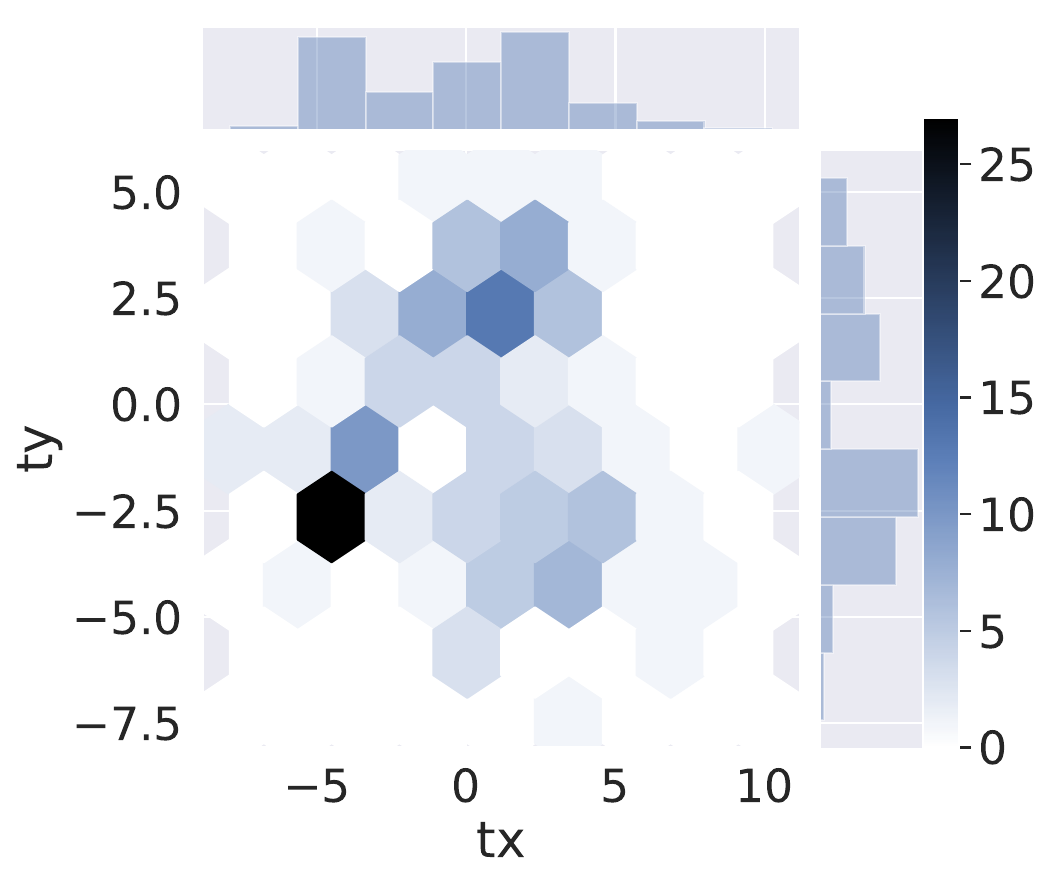}
         \caption{}
         \label{fig:shifts}
     \end{subfigure}
     \quad
     \begin{subfigure}[b]{0.20\textwidth}
         \centering
         \includegraphics[width=\textwidth]{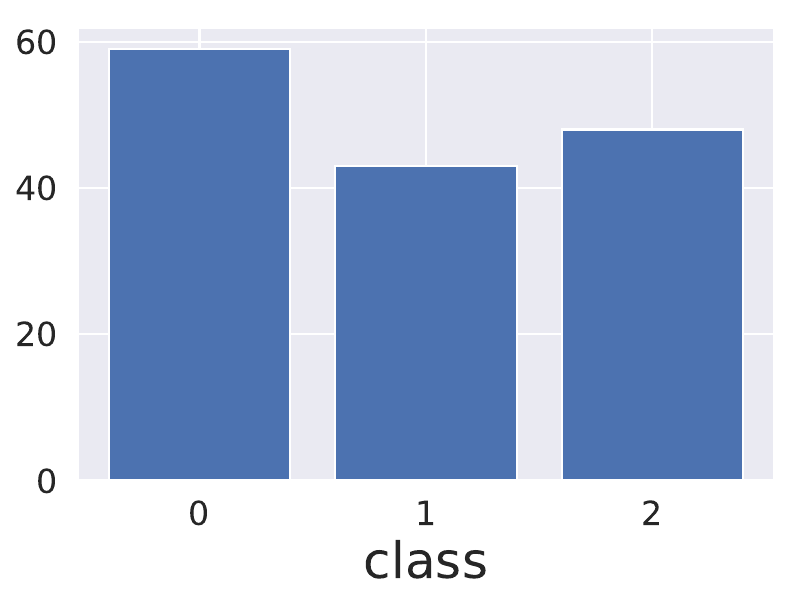}
         \caption{}
         \label{fig:class}
     \end{subfigure}
     }
     \mbox{
     \begin{subfigure}[b]{0.20\textwidth}
         \centering
         \includegraphics[width=\textwidth]{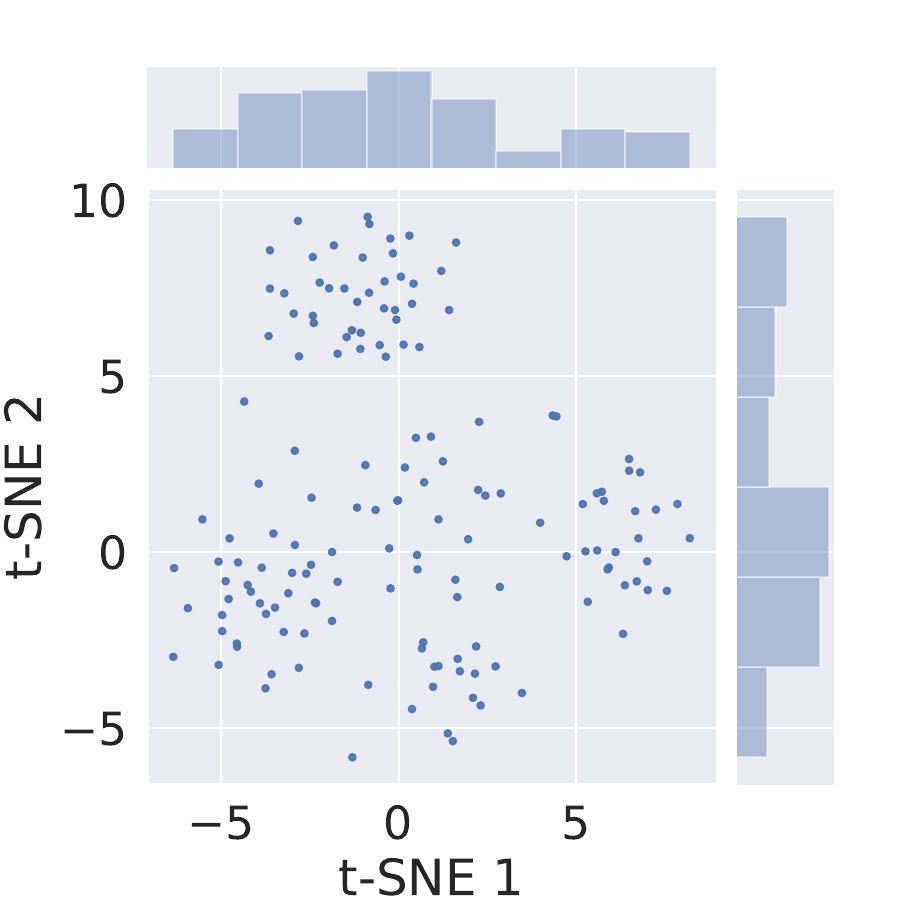}
         \caption{}
         \label{fig:tsne1}
     \end{subfigure}
     \quad
     \begin{subfigure}[b]{0.20\textwidth}
         \centering
         \includegraphics[width=\textwidth]{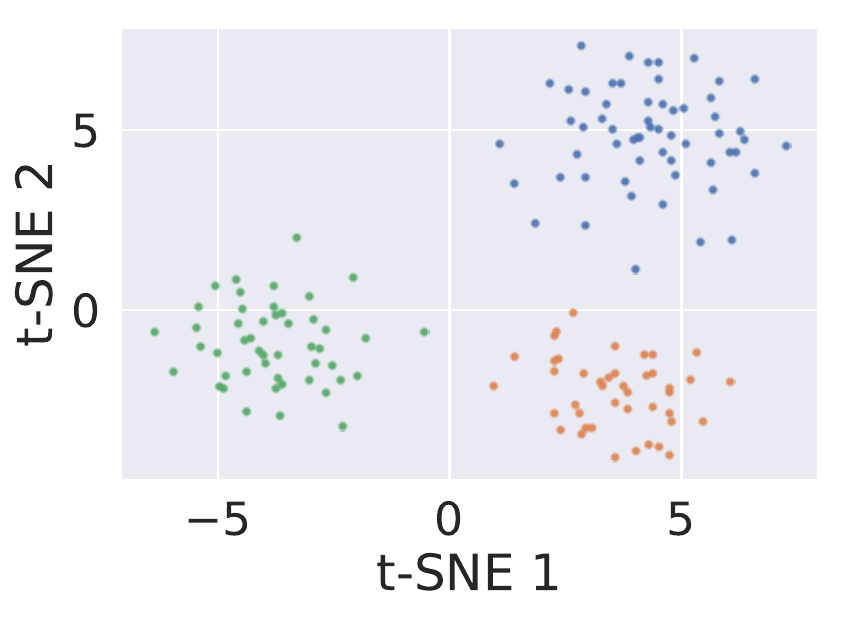}
         \caption{}
         \label{fig:tsne3}
     \end{subfigure}
     }
        \caption{Histogram of (a) x and y shift, (b) class assignments. t-SNE plot of (c) all particles before alignment, (d) after alignment with each class labeled in different colors. From (c),(d), we can perceive that the particles are better aligned after MRA as it groups nearest neighbors into the same class.}
        \label{fig:notebook}
\end{figure}

The parallel reduction scheme we employed to find the maximum value of each row in the table is shown in Fig. \ref{fig:reduction}. Here, we are not deciding how many threads to launch based on the data size since the length of each row is very large and can easily exceed the maximum number of threads allowed in a block. Instead, we launch a fixed number of threads and let each thread loop through memory, computing partial reduction operations on each element to initialize the memory. 
Finally, the standard parallel reduction is employed to reduce the initialized memory to find the largest element and its index. It is noted that our approach also enables coalesced access of the memory within each block during the initialization which can improve the memory access speed. After finding the index for each image, the corresponding alignment parameters are calculated for each image in parallel with GPU kernel. Finally, the new class averages are calculated through CuPy \cite{cupy_learningsys2017} using our custom class depicted in next section. 

\section{Proposed Software Stack}
The software stack of our library is shown in Fig. \ref{fig:soft_stack}. Popular computational cryo-EM library are built around C and C++ \cite{Relion,tang2007eman2}. On the other hand, the recent trend of packages designed for data analysis and data visualization is to build upon Python libraries. However, there is no convenient way to perform cryo-EM data analysis and visualization in the Python environment.  In our framework, the functionality of EMAN2 and the CUDA code is made available through several well-defined interfaces. We used CuPy to implement the required interoperability layer. To compute the averages and calculate the statistics of the transformed images, we implement a custom class as the GPU array interface. This class encapsulates the \textit{\cdbackslash cuda\_array\_interface\cdbackslash} which is created for interoperability between different GPU arrays in various python projects. With this class, we can easily access the GPU buffer of the transformed images and calculate them in Python to reduce the need to transfer data between host and device as much as possible. 
In the Python environment, we use NumPy \cite{harris2020array} and CuPy array to represent the particle images; it can be transformed into other popular computer vision or machine learning libraries for analysis \cite{scikit-learn}. Finally, the metadata is represented as Pandas \cite{reback2020pandas} dataframe for exploratory data analysis. The images and metadata are stored into either HDF5 or Berkeley DB (BDB) files. 
Developer or user can use the Jupyter notebook interface with our API as a pipeline for common data analysis. Several notebooks are included in the library to show that basic image operations like rotation and shift can be easily accelerated and the data analysis/visualization can be performed on the platform. 

The processing pipeline of our framework is illustrated as follows. First, the data source (cryo-EM images and metadata stores in HDF5, BDB or text file) are to be read in through the Python wrapper in parallel. Second, the user performs preprocess or exploratory data analysis in the Jupyter notebook. Third, the alignment and clustering are performed by calling the MRA script. Finally, the transformed images and metadata can be read into a notebook for further analysis or visualization. Fig. \ref{fig:notebook} shows an example of exploratory data analysis. With the plotted 1D and 2D histogram of shifts and class assignments from MRA procedure are plotted, we can then analyze the performance on particle picking or diagnose the problems like preferred orientations or the attraction by large clusters \cite{sorzano2010clustering}. In addition, we use 2SDR \cite{chung2020two} and t-SNE \cite{maaten2008visualizing} to plot the 2D embedding of all particles as shown in Fig. \ref{fig:notebook} (c),(d). We can then get an idea about the alignment progress and the performance of 2D classification algorithm.

%


\begin{figure}[htb]
\centering
\centering
 \mbox{
 \begin{subfigure}[b]{0.20\textwidth}
     \centering
     \includegraphics[width=\textwidth]{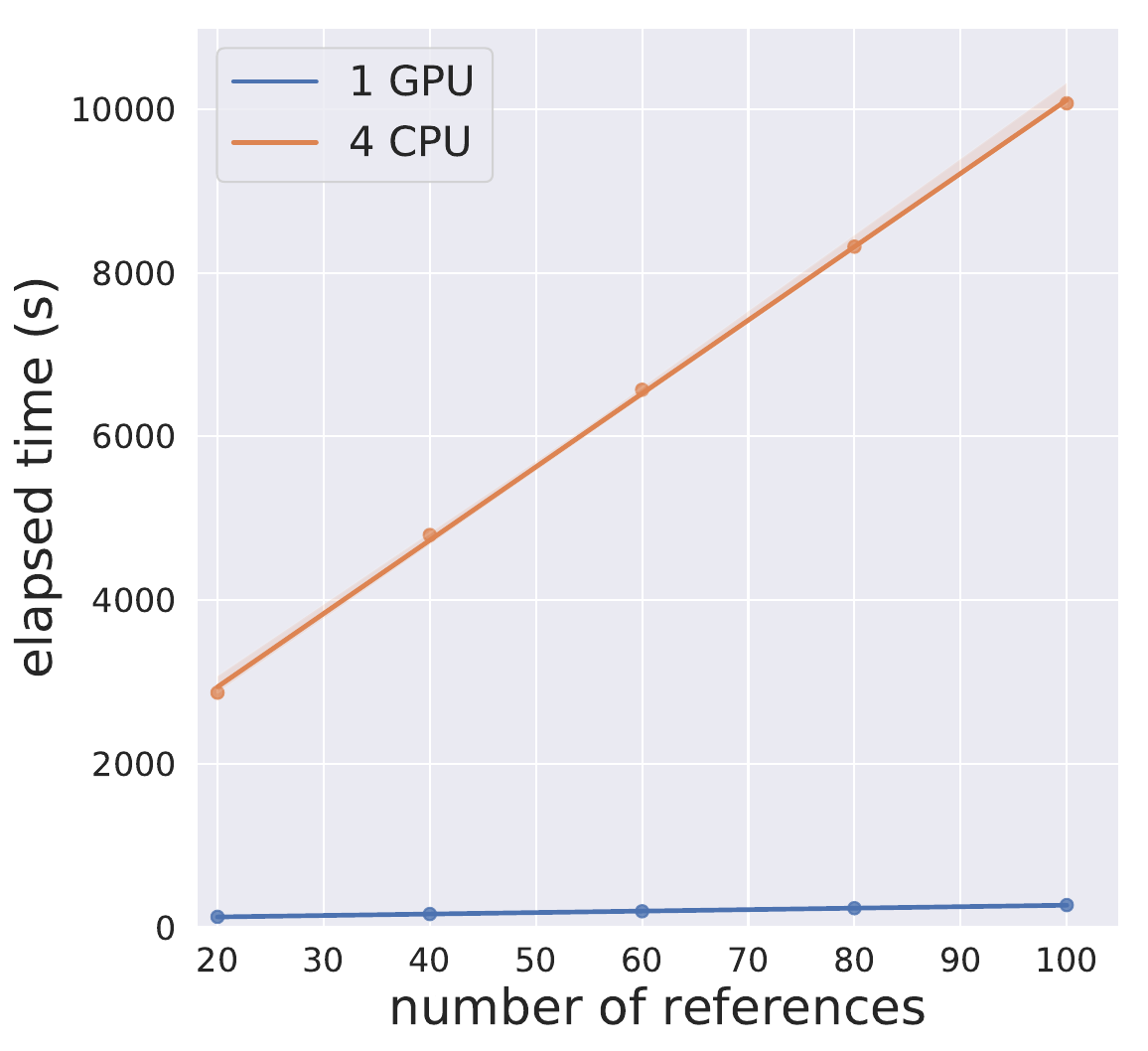} 
     \caption{}
     \label{fig:mr_time}
 \end{subfigure}
\quad
 \begin{subfigure}[b]{0.20\textwidth}
 \centering
\includegraphics[width=\textwidth]{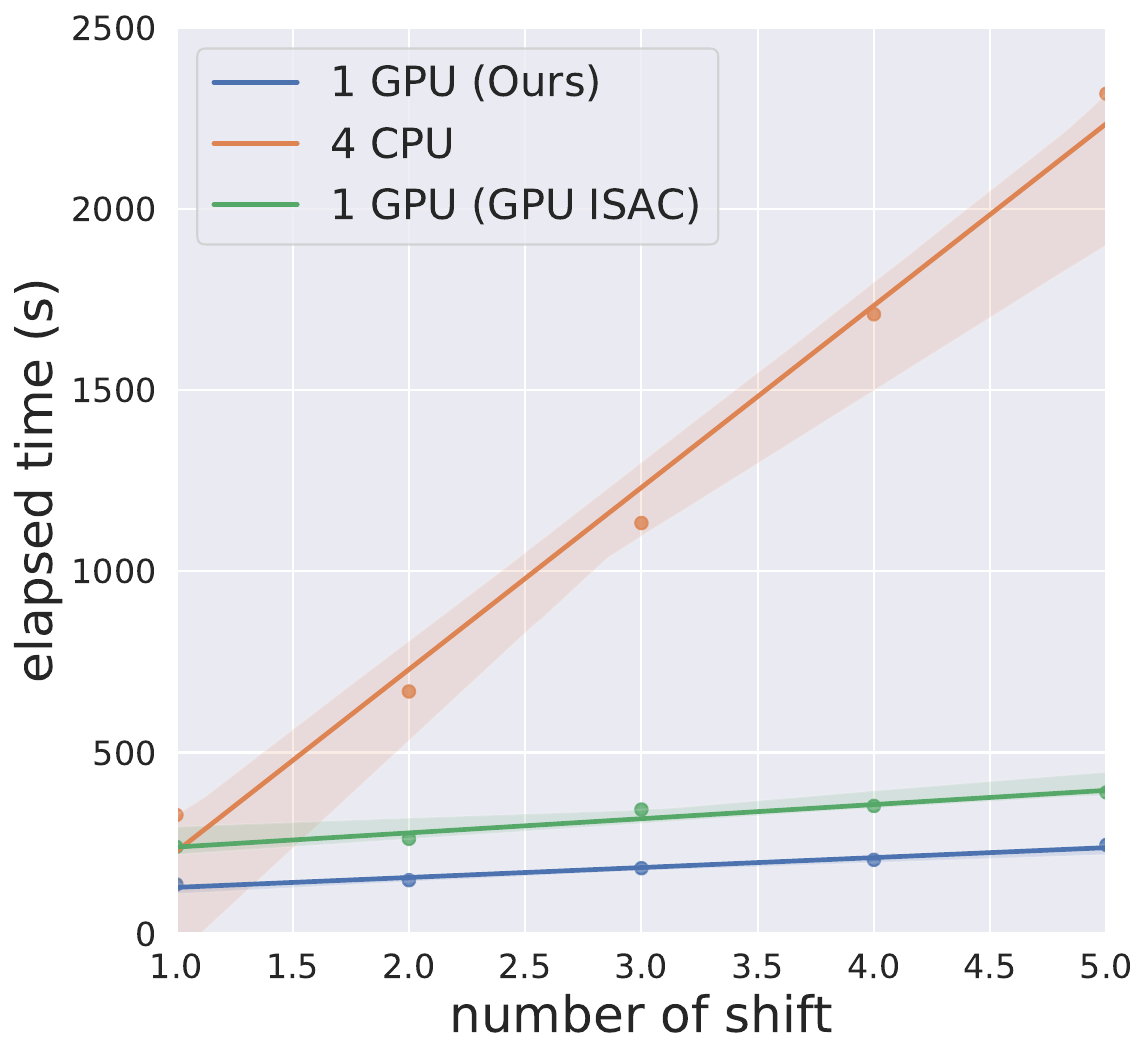} 
    \caption{}
     \label{fig:rf_time}
 \end{subfigure}
 }
 \mbox{
 \begin{subfigure}[b]{0.20\textwidth}
 \centering
\includegraphics[width=\textwidth]{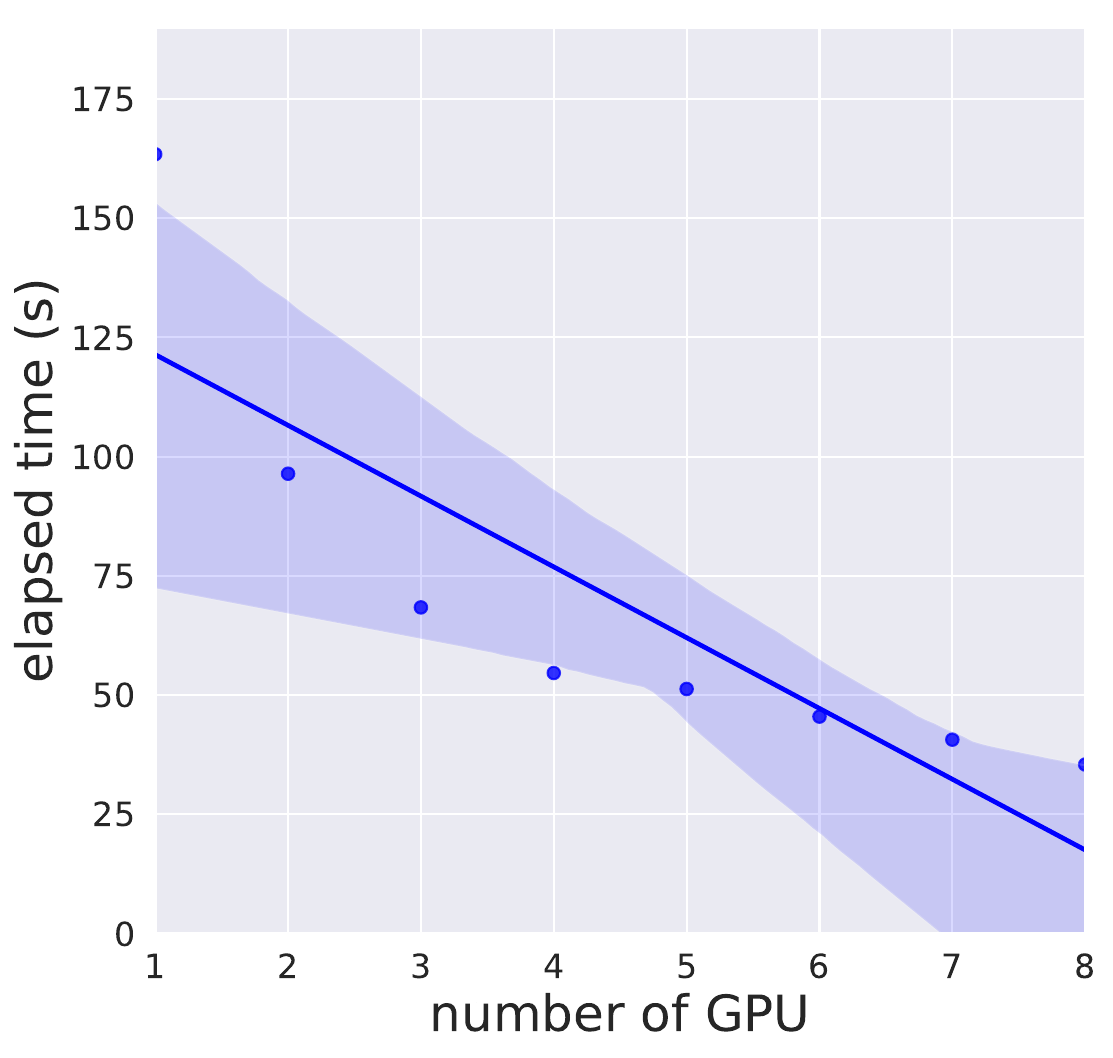} 
    \caption{}
     \label{fig:mr_time2}
 \end{subfigure}
 \quad
 \begin{subfigure}[b]{0.20\textwidth}
 \centering
\includegraphics[width=\textwidth]{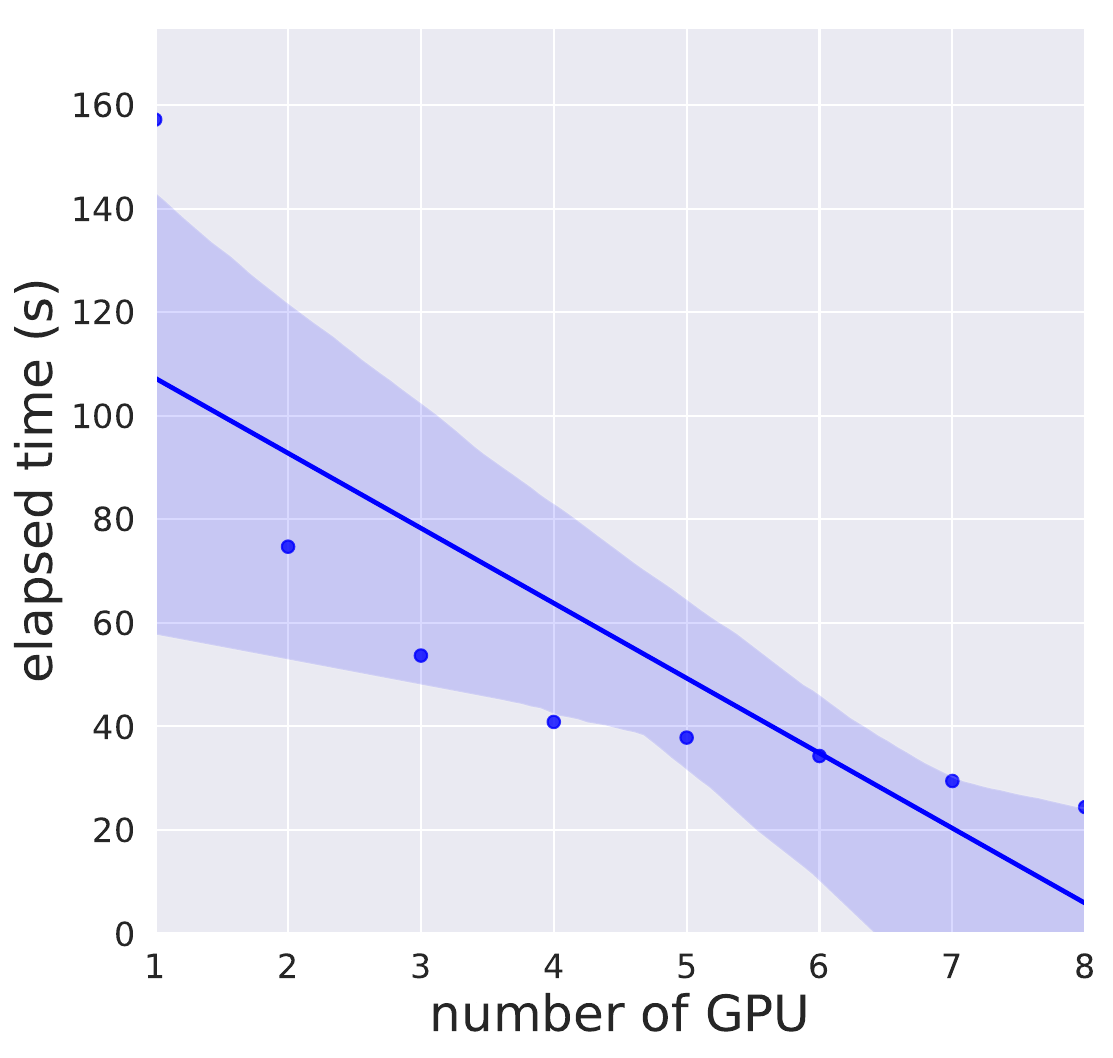} 
    \caption{}
     \label{fig:rf_time2}
 \end{subfigure}
}
\caption{(a) Computation time of MRA for a different number of references. (b) Computation time of RFA for different search ranges of shift. (c) The scaling behavior of MRA. (d) The scaling behavior of RFA.}
\label{fig:time}
\end{figure}

\section{The Experiment Results}
Comparisons between different implementations are conducted on TaiWan Computing Cloud (TWCC). The programs were running in the TWCC c.super instance, with 4 cores of Xeon Gold 6154 processors and an Nvidia Tesla V100 GPU. The benchmark dataset of Ribosome 80s \cite{wong2014cryo} from RELION is used with height and width both down-sampled to 90 pixels. We first compared the CPU implementation of MRA from EMAN2 as shown in Fig. \ref{fig:time}(a). The search range in the x and y direction is set to 3 and the radius of particles is 36 pixels. We can perceive that the speedup is $22\times$ to $37\times$ with different reference numbers. We then compared the computation time of CPU implementation of RFA from EMAN2 as shown in Fig. \ref{fig:time}(b). The search step is set to 1 and the speedup is $2.4\times$ to $9.4\times$ with different 2D shifts. We also compared with the RFA in GPU ISAC and the speedup is about $1.6\times$. Fig. \ref{fig:time}(a) presents the performance of our implementation increase that is gained over multiple-GPU on the c.4xsuper instance with 8 GPUs. 
It can be observed that our implementation scales well from 1 to 4 GPUs and provides near linear speedup compared to single GPU. On the other hand, when the number of GPU is greater than 4 the speedup is saturated because the time is dominated by the data transfer.

\section{Conclusion and Future Works}
With continuing enhancement of processing algorithms, cryo-EM has become  a progressively powerful and efficient technique to solve structures of molecules. Currently, the
performance of the popular GPU-accelerated Bayesian approach for 2D Classification in
cryo-EM still poses a bottleneck in the processing as the computation time and
the classification quality are sub-optimal. As a result, 
MRA-based approaches represent an option as it can better reveal structural variations at the 2D level. To the best of our knowledge, until now MRA-based methods are mostly implemented with CPU-supported algorithms and thereby time-consuming. In this work, a library termed \textit{Cryo-RALib} that expands the functionality of the CUDA library used by GPU ISAC is introduced. The library contains a GPU-accelerated MRA routine that can be used to accelerate 2D classification algorithms. We also provide several well-defined interfaces to connect the cryo-EM image analysis with the python data science stack to help users to perform analysis and visualization more easily. Our benchmarking results on TWCC show this implementation can accelerate the alignment procedure by one order of magnitude. In the future, it would be interesting to expand the library to accommodate other cryo-EM packages into this framework, and to provide a friendly environment for users and developers as well.

\section*{Acknowledgement}
The authors gratefully acknowledge Ryan Jeng from Nvidia and the NCHC GPU Hackathon for the help on our research project. The authors also acknowledge the open-source projects that we built upon, especially the EMAN2 and GPU ISAC 2.3.2. Finally, the authors acknowledge the insightful suggestions from Dr. Stefan Raunser.

%
%
\bibliographystyle{IEEEbib}
\bibliography{ref}

\end{document}